УДК 004.738.2

А.О. Дадукин, Н.И. Пчелинцева

# ТЕХНОЛОГИЯ СОЗДАНИЯ ПОРТАТИВНЫХ УСТРОЙСТВ МЕДИЦИНСКОЙ НАПРАВЛЕННОСТИ С ИСПОЛЬЗОВАНИЕМ МОДУЛЯ BLUETOOTH

*Статья посвящена спецификации беспроводных персональных сетей Bluetooth, обеспечивающая обмен информацией между такими устройствами, как персональные компьютеры, мобильные телефоны, принтеры, цифровые фотоаппараты, мышки, клавиатуры, джойстики, наушники, гарнитуры на надёжной, бесплатной, повсеместно доступной радиочастоте для ближней связи. Показано развитие технологии и её применение в проектировании устройств. Подробно рассмотрен Health Device Profile, главной особенностью которого является работа с устройствами медицинской направленности.*

***Ключевые слова:*** *протоколы передачи данных, модуль Bluetooth, устройства медицинской направленности, пикосети.*

**Введение.** В настоящее время в связи со стремительным развитием информационных технологий, более доступным и перспективным направлением является создание портативных умных устройств. Увеличение доступности разработки позволило многим компаниям сосредоточить свое внимание на наиболее актуальных проблемах человечества. Одной из таких проблем, безусловно, является применение портативных устройств в области медицинских исследований.

**Аппаратное обеспечение портативных устройств.** С учетом всех недостатков [1] подобное устройство может выглядеть следующим образом (Рис. 1):

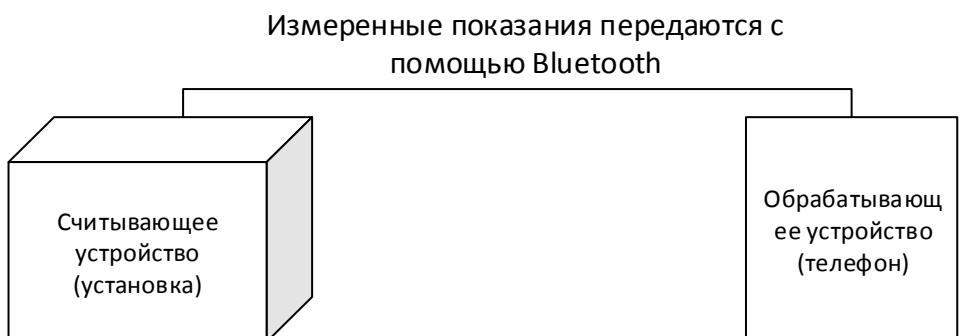

**Рис. 1.** Архитектура портативного устройства

Необходима установка, позволяющая считывать биометрические показатели человека (частота сердечных сокращений, длительность наполнения, индекс восходящей волны) и отсылать полученные данные на обрабатывающее устройство. Данные с устройства считывания передаются при помощи спецификации Bluetooth Health Profile, подразумевающий наличие клиент-серверной архитектуры системы во время передачи данных. Одно из устройств – считывающая установка – будет являться сервером, поддерживающим постоянное соединение через Bluetooth-сокет, другое – телефон – Bluetooth –клиентом (Рис. 2) [2].

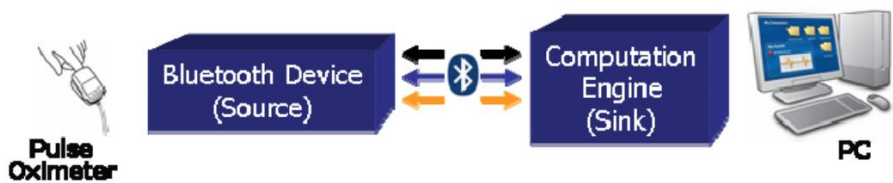

**Рис. 2.** Схема работы пульсометра

Первое, что необходимо определить при проектировании подобного рода систем – это аппаратная архитектура. С учетом всех вышеизложенных недостатков и требований необходимо дифференцировать этапы разработки на независимые итерации. Первой итерацией данного цикла будет являться – создание рабочего прототипа. Для разработки первой рабочей пары была выбрана независимая платформа Arduino и устройство под управлением Android OS. Выбор основан на относительной дешевизне модулей и скоростью разработки под данные платформы.

Система, состоящая из обособленного устройства и мобильного телефона, должна поддерживать постоянное соединение для создания условий общения в режиме реального времени двух независимых платформ. Так как устройство подразумевает модульность и относительную независимость программных частей, то необходимым требованием к установке является OTA взаимодействие.

При выборе протокола общения необходимо руководствоваться следующими соображениями:
- ♦ модуль, осуществляющий общение должен быть установлен в подавляющее большинство устройств;
- ♦ реализуемый протокол должен быть открытым;
- ♦ модуль не требует работы на значительных расстояниях, следовательно, высокая энергоемкость.
- ♦ Исходя из этих требований, наиболее подходящей технологией будет – Bluetooth.

**Технология Bluetooth.** *Bluetooth* – международный стандарт беспроводных коммуникаций малого радиуса действия. Основная область применения – обеспечение экономичной радиосвязи между различными электронными устройствами [3]. Основной акцент ставится на малую габаритность данного модуля, долгое время автономной работы и относительную дешевизну.

Сама технология существует довольно длительное время и пришла на замену технологии RS-232 для передачи данных. У истоков данной разработки стояла компания Ericsson, в дальнейшем к ее разработки присоединились такие крупные компании как Nokia, Toshiba и другие. В результате в 1999 году был образован комитет по стандартизации технологии Bluetooth - Special Interest Group.

Для работы данной технологии используется нижний диапазон ISM (Industrial, Scientific, Medical) [4], предназначенный для работы промышленных устройств, устройств научной и медицинской направленности. Основным преимуществом данного диапазона является то, что он свободен от лицензирования практически во всех странах мира (в том числе и в России), это значит, что для работы сертифицированных устройств не требуется дополнительное разрешение.

Изначально, все Bluetooth-устройства находятся в режиме *ожидания* (*Standby*). В данном режиме устройство сканирует 32 фиксированные частоты с интервалом - раз в 1.28 секунды.

При приеме запроса устройство реагирует на него в соответсвии с установленным режимом. При этом возможны следующие способы реакции на запрос:

♦ устройство находится в режиме отклика (*discoverable mode*), при этом оно всегда отвечает на полученные запросы;
♦ устройство находится в режиме частичного (ограниченного) отклика (*limited discoverable mode*), при этом оно отвечает только ограниченного время (в соответсвии с выставленной продолжительностью) или ограничивается другими факторами;
♦ устройство находится в режиме отказа отклика (*non-discoverable mode*) и не принимает любые входящие устройства.

Кроме данных трех режимов само устройство может быть подключаемым или недоступным (connectable/non-connectable mode). Если устройство недоступно оно не позволяет проводить сканирование и ограничивает передачу данных.

После того как устройства успешного подключились друг к другу, они определяют диапазон частот, с которым им предстоит работать, порядок и частоту изменения рабочего диапазона, размер страниц и некоторые другие параметры соединения.

Каждое Bluetooth устройство имеет свой уникальный глобальный адрес, однако на уровне взаимодействия с пользователем это может быть и имя Bluetooth-устройства, причем не обязательно уникальное.

Все Bluetooth установки могут устанавливать многоточечные соединения. При этом они объединяются в пикосети (Рис. 3).

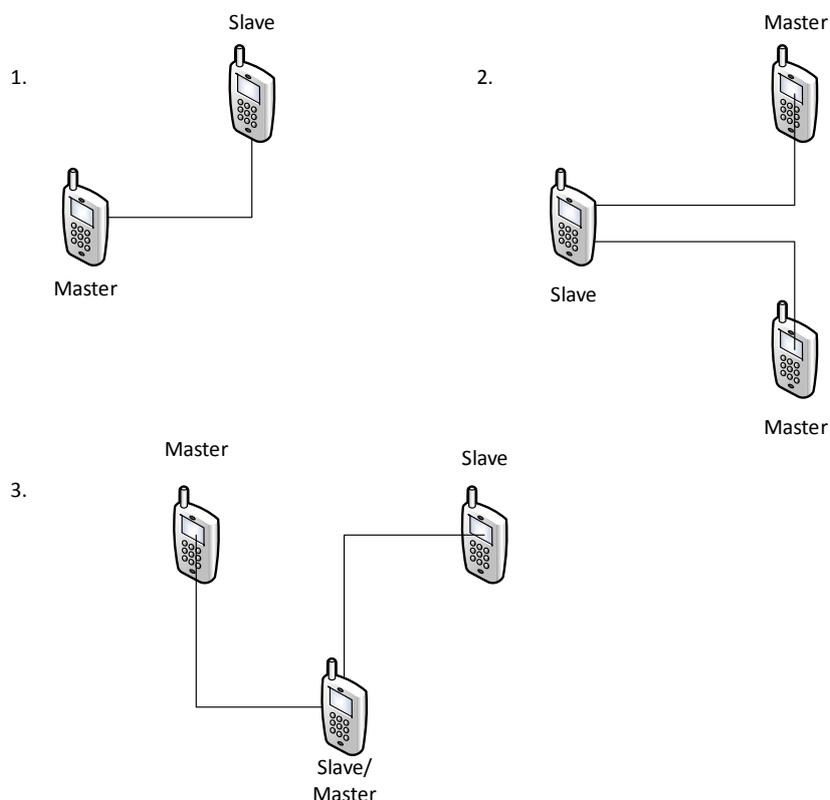

**Рис. 3.** Элементарная пикосеть

Элементарная пикосеть – это два устройства с модулями Bluetooth, которые называются master (главный) и slave (ведомый). Причем master выполняет инициацию и поддержку функционирования соединения. Максимальное количество соединений для одного мастера – 7, а суммарная скорость передачи данных не превышает максимум для данной версии технологии. Функции master и slave жестко не фиксируются за устройствами и в зависимости от загрузки могут меняться. Причем, в зависимости от структуры пикосети устройство в разных соединениях может выполнять разные роли, а также может быть в качестве slave для разных master.

Преимущества Bluetooth по сравнению с другими способами передачи данных (кабель, IRDA (инфракрасный порт)) очевидно.

Кабель – классический способ соединения. Он обеспечивает надежную связь между телефоном и компьютером, но у него есть минус, который в большой степени нивелирует его достоинства, – это неудобство использования. Вы очень сильно ограничены в своем передвижении длиной кабеля.

IRDA решает эти проблемы – нет соединительного провода между телефоном и компьютером. Но существенным минусом IRDA является природа инфракрасного излучения, как волн в световом диапазоне. Это обстоятельство налагает необходимость прямой видимости для связи между двумя устройствами.

**Безопасность Bluetooth.** Безопасность в Bluetooth соединении обеспечивается реализацией алгоритма аутентификации и генерации ключа SAFER+. Генерация инициализационного и главного ключа осуществляется по алгоритму E22. Поточный шифр E0 используется для закрытия передаваемых данных.

**Профили Bluetooth.** Группа разработки Bluetooth SIG определила и одобрила несколько десятков разных профилей, ниже приведены некоторые из них [3]:

*Таблица 1*

| Профиль | Назначение |
|---|---|
| Audio / Video Remote Control Profile (AVRCP) | A2DP создан для передачи двухканальных стерео аудиоданных, например, музыки, между Bluetooth устройствами. В профиле полностью поддерживается низкосжатый кодек Sub_Band_Codec (SBC) и может поддерживать MPEG-1,2 аудио, MPEG-2,4 AAC и ATRAC, а также кодеки определенные производителем. |
| Health Device Profile (HDP) | Профиль разработан для работы сертифицированных Healthcare и Fitness устройств. |
| Human Interface Device Profile (HID) | Профиль разработан для соединения с устройствами HID (Human Interface Device): мышки, джойстики, клавиатуры и проч. Использует медленный канал, работает на малой мощности. |
| Headset Profile (HSP) | Профиль обеспечивает соединение беспроводной гарнитуры и телефона. Используется стандартный набор AT команд спецификации GSM 07.07, что позволяет делать звонки, отвечать на звонки, заканчивать звонок, управлять громкостю. |
| SIM Access Profile (SAP, SIM) | Профиль используется для обеспечения доступа к SIM-карте телефона, что дает возможность использовать одну SIM-карту для нескольких устройств. |
| Synchronisation Profile (SYNCH) | Профиль используется для синхронизации личных данных (PIM). |
| Video Distribution Profile (VDP) | Профиль разработан для передачи потокового видео. Поддерживает стандарт H.263, могут поддерживаться MPEG-4 Visual Simple Profile, H.263 profiles 3, profile 8. |
| Wireless Application Protocol Bearer (WAPB) | Протокол обеспечивает организацию PPP (Point-to-Point) соединения. |

**Health Device Profile.** Данный профиль определяет требования для сертифицированных Bluetooth Healthcare и Fitness устройств (относящихся к группе 'health') [4]. Профиль HDP используется для подключения устройств-источников таких как: мониторы артериального давления, весы, глюкометры, термометры, пульсоксиметры для передачи данных на мобильный телефоны, ноутбуки, а также стационарные компьютеры. Это в свою очередь делает возможным соединение без кабеля для устройств медицинской направленности. Данный профиль базируется MCAP и некоторые функции L2CAP, например, режим потоковой передачи [6].

Эта версия HDP, основанная на MCAP, предусматривает следующее:

- ♦ Дополнительная синхронизация таймеров с микросекундой точностью для спаренных устройств, позволяющая сравнивать временные метки напрямую
- ♦ Определяет эффективный метод восстановления соединения для большей экономии электроэнергии

HDP – специализированный профиль для приложений в области здравоохранения, поэтому обладает некоторыми преимуществами по сравнению с другими видами профилей:

- ♦ Берет на себя задачи по проверки требований, подключаемых устройств на совместимость [5].
- ♦ Обеспечивает максимальную совместимость устройств, работающих по ISO/IEEE.
- ♦ 11073-20601 Personal Health Data Exchange Protocol, который определяет протокол обмена данными Data Exchange Protocol.
- ♦ Ориентирован на поддержание соединения, чтобы обеспечить более устойчивое поведение в случаях, когда источник находится в движении, перемещается к границе диапазона, или отсоединяется (случайно или намеренно), позволяя принимающей стороне своевременно принять необходимые меры.
- ♦ Позволяет обслуживать одновременно несколько каналов передачи данных.
- ♦ Относительно недорогой в реализации и нетребовательный в исполнении.

Однако, несмотря на все возможность HDP он не предназначен для передачи звука (в том числе и медицинского характера), например, со стетоскопа [2].

**Заключение.** Перспективы дальнейшего исследования проблемы мы видим в более детальном изучении работы данного протокола и создании на основе него рабочей пары (прототипа), способного передавать большие серии измеренных данных.

**СВЕДЕНИЯ ОБ АВТОРАХ**

**Дадукин Александр Олегович** (Dadukin A.O.) - студент КФ МГТУ им. Н.Э. Баумана; *alexanderdadukin@gmail.com*

**Пчелинцева Наталья Ибрагимовна** (Pchelintseva N.I.) - канд. техн. наук, доцент КФ МГТУ им. Н.Э. Баумана; *pchelintseva.n@yandex.ru*

**PORTABLE MEDICAL DEVICES CREATION TECHNOLOGY BY USING THE BLUETOOTH MODULE**

*The article is devoted Bluetooth wireless personal area networks specification, which provides standard for exchanging data over short distances. It is shown how the technology has evolved and its application in the design of devices. Health Device Profile considered in details, which the main feature is the work of a medical orientation devices.*

***Keywords:*** *data transfer protocols, Bluetooth module, medical devices, piconet.*